\PassOptionsToPackage{dvipsnames,usenames}{xcolor} 

\documentclass[journal=nalefd,manuscript=letter,articletitle=false,layout=twocolumn]{achemso}

\usepackage[colorlinks,pdfencoding=auto,psdextra]{hyperref}
\usepackage{doi} 

\usepackage[T1]{fontenc} 
\usepackage{amsmath} 
\usepackage{amssymb} 
\usepackage{placeins}
\usepackage[usenames, dvipsnames]{xcolor} 
\usepackage{ulem} 
\usepackage{orcidlink}

\setkeys{acs}{doi = true}

\author{Erik Kirstein~\orcidlink{0000-0002-2549-2115}}
\affiliation{Authors contributed equally}
\alsoaffiliation{Experimental Physics 2, Department of Physics, TU Dortmund,  44227 Dortmund, Germany}
\email{erik.kirstein@tu-dortmund.de}

\author{Evgeny A. Zhukov~\orcidlink{0000-0003-0695-0093}}
\affiliation{Authors contributed equally}
\alsoaffiliation{Experimental Physics 2, Department of Physics, TU Dortmund, 44227 Dortmund, Germany}
\alsoaffiliation{Ioffe Institute, Russian Academy of Sciences, 194021 St. Petersburg, Russia}

\author{Dmitri R. Yakovlev~\orcidlink{0000-0001-7349-2745}}
\affiliation{Experimental Physics 2, Department of Physics, TU Dortmund, 44227 Dortmund, Germany} 
\alsoaffiliation{Ioffe Institute, Russian Academy of Sciences, 194021 St. Petersburg, Russia}
\email{dmitri.yakovlev@tu-dortmund.de}

\author{Nataliia E. Kopteva~\orcidlink{0000-0003-0865-0393}}
\affiliation{Experimental Physics 2, Department of Physics, TU Dortmund, 44227 Dortmund, Germany} 

\author{Carolin Harkort~\orcidlink{0000-0003-1975-9773}}
\affiliation{Experimental Physics 2, Department of Physics, TU Dortmund, 44227 Dortmund, Germany}

\author{Dennis Kudlacik~\orcidlink{0000-0001-5473-8383}}
\affiliation{Experimental Physics 2, Department of Physics, TU Dortmund, 44227 Dortmund, Germany}

\author{Oleh Hordiichuk~\orcidlink{0000-0001-7679-4423}}
\affiliation{Department of Chemistry and Applied Biosciences, ETH Z\"urich,  Z\"urich CH-8093, Switzerland}
\alsoaffiliation{EMPA-Swiss Federal Laboratories for Materials Science and Technology, D\"ubendorf CH-8600, Switzerland}

\author{Maksym V. Kovalenko~\orcidlink{0000-0002-6396-8938}}
\affiliation{Department of Chemistry and Applied Biosciences, ETH Z\"urich, Z\"urich CH-8093, Switzerland}
\alsoaffiliation{EMPA-Swiss Federal Laboratories for Materials Science and Technology,  D\"ubendorf CH-8600, Switzerland}

\author{Manfred Bayer~\orcidlink{0000-0002-0893-5949}}
\affiliation{Experimental Physics 2, Department of Physics, TU Dortmund, 44227 Dortmund, Germany} 

\abbreviations{IR,NMR,UV}

\makeatletter
\title[PePI Spin Dynamics] {Coherent Spin Dynamics of Electrons in Two-Dimensional (PEA)$_2$PbI$_4$ Perovskites}
\let\Title\@title
\makeatother

\begin{document}


\textbf{Abstract}

The versatile potential of lead halide perovskites and two-dimensional materials is merged in the Ruddlesen-Popper perovskites having outstanding optical properties. Here, the coherent spin dynamics in Ruddlesen-Popper (PEA)$_2$PbI$_4$ perovskites are investigated by picosecond pump-probe Kerr rotation in an external magnetic field. The Larmor spin precession of resident electrons with a spin dephasing time of 190~ps is identified. The longitudinal spin relaxation time in weak magnetic fields measured by the spin inertia method is as long as 25~$\mu$s.  A significant anisotropy of the electron $g$-factor with the in-plane value of $+2.45$ and out-of-plane value of $+2.05$ is found. The exciton out-of-plane $g$-factor is measured to be of $+1.6$ by magneto-reflectivity. This work contributes to the understanding of the spin-dependent properties of two-dimensional perovskites and their spin dynamics.


\clearpage

The first reports on optical studies of excitons in Ruddlesden-Popper type perovskite structures $A$PbI$_4$, where $A$ is an organic anion, date back to 1989. \cite{ishihara1989,ishihara1990,kataoka1993}. These crystals contain monolayers of the [PbI$_6$] octahedra, which form quantum wells, separated by organic barriers. In these two-dimensional (2D) structures the exciton binding energy is greatly increased up to $200-500$~meV due to the reduction of dimensionality and the dielectric confinement \cite{muljarov1995}. In this respect, they are similar to 2D materials like the transition metal dichalcogenides~\cite{blancon2020,wang2018}. Recently perovskite materials have regained strong interest due to the great progress of lead halide perovskites, and have advanced as versatile platform for optoelectronic applications~\cite{mao2018,lempicka-mirek2022,hoye2022,vardeny2022c,vardeny2022d} and beyond~\cite{ricciardulli2021}. In photovoltaics a power conversion efficiency above 17\% \cite{sidhik2021,seitz2020} was reached and long term stability was reported~\cite{grancini2017}. 

A variety of optical methods, like four-wave mixing \cite{thouin2018}, spatially resolved spectroscopy \cite{seitz2020}, second harmonic generation \cite{qin2022}, strong coupling in cavities \cite{lempicka-mirek2022}, etc. were used to address the optical properties of exciton complexes in 2D perovskites.
The application of high magnetic fields up to 60~T delivered information about the exciton binding energy, effective mass, fine structure, and $g$-factor~\cite{blancon2018,dyksik2020,dyksik2021,do2020b,surrente2021}. Time-resolved transmission was used to study the exciton spin dynamics at picosecond and subpicosecond time scales~\cite{giovanni2018,chen2021,bourelle2022,bourelle2020}. However, in these reports, carried out at temperatures above liquid nitrogen temperature up to ambient conditions, the spin dynamics were significantly limited, and spin relaxation times exceeding a few ps have not been reported so far.  

The experimental techniques established in spin physics, like time-resolved Faraday/Kerr rotation were successfully used for investigation of lead halide perovskites bulk crystals \cite{belykh2019,kirstein2021,kirstein2022,kirstein2022b}, polycrystalline films\cite{odenthal2017,garcia-arellano2021,garcia-arellano2022}, and nanocrystals~\cite{crane2020,grigoryev2021,kirstein2022SML}. These techniques provide comprehensive information on the coherent and incoherent spin dynamics of electrons and holes, on their Land\'e factors ($g$-factors) and on the charge carrier interaction with the nuclear spin system. We are not aware of such experiments performed on the 2D perovskites. Note that the $g$-factor values and their anisotropy give access to the band structure parameters through comparison with the results of modeling  accounting also for the reduced dimensionality. 

In this paper, we use time-resolved Kerr rotation (TRKR) to study the coherent spin dynamics of electrons in films composed of the two-dimensional (PEA)$_2$PbI$_4$ perovskite. The spin dephasing and longitudinal spin relaxation times are measured. The electron $g$-factor shows a considerable anisotropy, with a larger value for the in-plane component. The exciton out-of-plane $g$-factor component is measured in magneto-reflectivity, which allows us to evaluate the hole $g$-factor in this direction. 


\subsection{Optical properties}

The (PEA)$_2$PbI$_4$ Ruddlesden-Popper type perovskite structure consists of a stack of corner shared PbI$_6$ octahedra monolayers separated by van der Waals-bonded pairs of PEA [phenethylammonium (C$_6$H$_5$)CH$_2$CH$_2$NH$_3$] molecules. In contrast to the $A$PbI$_3$ lead iodine archetype bulk crystal, with the $A$ cation being \{Cs, methylammonium (MA), or formamidinium (FA)\}, with the band gap energy in the near infrared spectral range around $1.7, 1.6,$ and $1.5$~eV, respectively~\cite{tao2019}, in the 2D (PEA)$_2$PbI$_4$ ($n=1$) the band gap is increased to about 2.6~eV by the quantum confinement~\cite{dyksik2020,blancon2018}. The exciton binding energy also strongly increases from about $15$~meV \cite{baranowski2020b} in bulk up to 260~meV~\cite{dyksik2020}. This increase is provided by the reduction of dimensionality and largely also by the dielectric confinement effect caused by the contrast in dielectric constants of the perovskite monolayers and the organic environment~\cite{blancon2018,straus2018,stier2016,baranowski2020,shornikova2021}.

\begin{figure*}[!ht]
  \includegraphics[width=\textwidth]{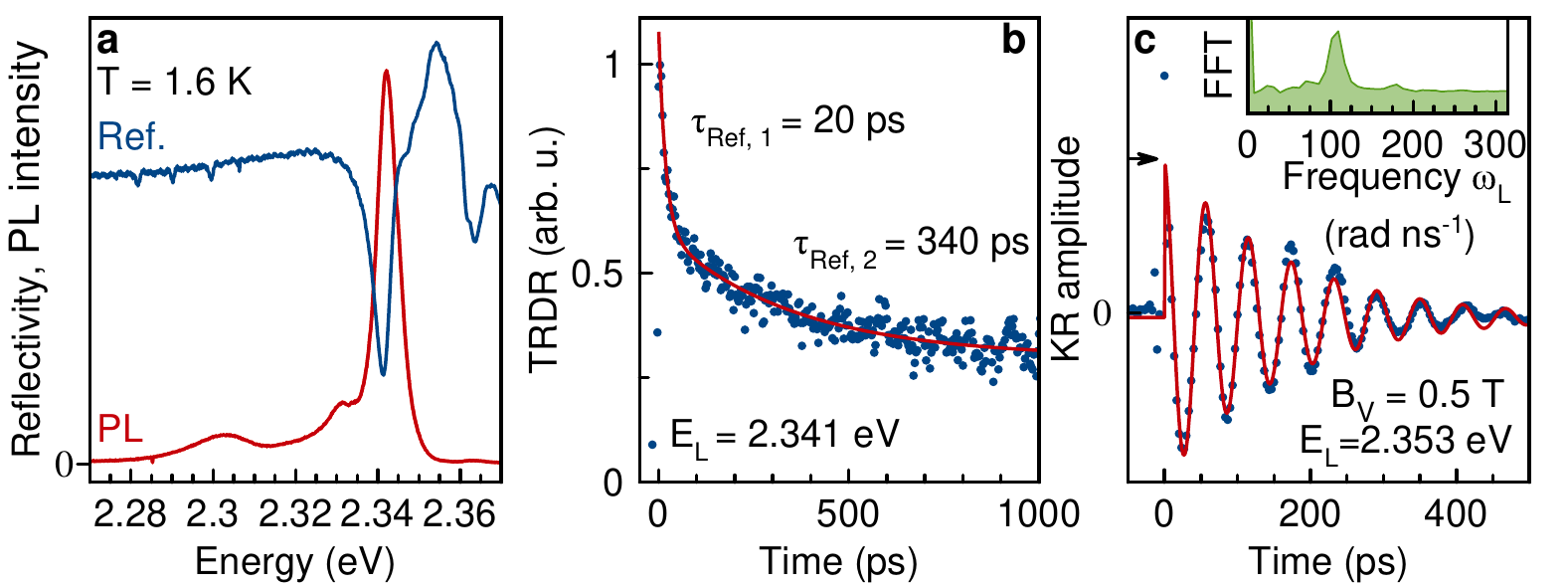}
\caption{(a) Reflectivity (blue) and photoluminescence (red) spectra of (PEA)$_2$PbI$_4$ measured at $T=1.6$\,K. 
(b) Time-resolved differential reflection dynamics (dots) measured with degenerate pump and probe energies at $E_\textrm{L}=2.341$\,eV for $T=1.6$\,K. Red line is a biexponential fit with the decay time constants $\tau_{\textrm{Ref,} 1}$ and $\tau_{\textrm{Ref,} 2}$. 
(c) Time-resolved Kerr rotation dynamics (dots) measured in Voigt geometry at $B_{\rm V} = 0.5$\,T for a laser energy of $E_\textrm{L}=2.353$\,eV, $P=1.5$\,mW. $T=1.6$\,K. Line is a fit with Eq.~\eqref{eq:KerrRot}. The arrow marks the KR amplitude $S$ at zero time delay. Inset gives the Fast Fourier Transform of the dynamics from the main panel.}
  \label{fig:Intro}
\end{figure*}

The studied (PEA)$_2$PbI$_4$ sample shows a pronounced exciton resonance in the reflectivity spectrum, measured at the temperature of $T=1.6$~K, Figure~\ref{fig:Intro}a. Its minimum is located at 2.341~eV and has a full width at half maximum of 6.6~meV. The photoluminescence spectrum (PL) shows a strong emission line with a similar width, and the maximum at 2.342~eV. It can be assigned to exciton emission of both free and weakly localized excitons. The weaker PL line at 2.330~eV can be assigned to dark exciton emission~\cite{neumann2021}, and the line at $2.303$~eV was associated in literature with bound excitons~\cite{neumann2021}, biexcitons~\cite{fang2020}, and a Rashba split ground state~\cite{do2020,zhai2017}.

The exciton and charge carrier population dynamics were measured by means of time-resolved differential reflection (TRDR) at the energy of 2.341~eV, Figure~\ref{fig:Intro}b. The dynamics are given by a decay with two times, $\tau_{\textrm{Ref}, 1} = 20$\,ps and $\tau_{\textrm{Ref}, 2} = 340$\,ps. Additionally, a long-living component, which lasts beyond the 1\,ns time window, was identified in accordance with literature reports~\cite{kahmann2021}. We attribute the short dynamics of 20~ps to the lifetime of bright excitons, which have a large oscillator strength in 2D perovskites. The lifetime is limited by their radiative recombination and relaxation into dark exciton states. This interpretation is in agreement with literature data on the low temperature recombination dynamics in (PEA)$_2$PbI$_4$~\cite{fang2020,giovanni2018,liu2019b}. The longer dynamics o 340\,ps can be attributed to nongeminate recombination of charge carriers, and the times exceeding 1\,ns are associated with trapping-detrapping processes of charge carriers.

\subsection{Coherent spin dynamics of electrons}

We use the time-resolved Kerr rotation (TRKR) technique~\cite{yakovlevCh6,kirstein2021} to study the coherent spin dynamics in the 2D (PEA)$_2$PbI$_4$ perovskites. Most experiments are performed at the low temperature of $T=1.6$~K, where the spin relaxation mechanisms have reduced efficiency. A typical example of the KR dynamics measured in the magnetic field of $B_{\rm V}=0.5$~T applied in the Voigt geometry is shown in Figure~\ref{fig:Intro}c. The laser photon energy is set to the maximum of the KR amplitude at 2.353~eV, which is slightly shifted to higher energy relative to the exciton resonance. The spectral dependence of the KR amplitude is given in the Supporting Information, Figure~S4. The spin dynamics in Figure~\ref{fig:Intro}c show an oscillating behavior, the KR amplitude, $A_{\rm KR}(t)$, decays within 500~ps. It can be well fitted with 
\begin{equation} 
\label{eq:KerrRot}
A_{\rm KR}(t) = S \cos{(\omega_{\rm L}t)}\exp(-t/T_2^*)\,.
\end{equation} 
Here $S$ is the spin polarization at zero time delay, $\omega_\textrm{L} = |g| \mu_\textrm{B} B/\hbar$ the Larmor precession frequency, $g$ the Land\'e factor ($g$-factor), $\mu_\textrm{B}$ the Bohr magneton, $T_2^*$ the spin dephasing time, $t$ the time delay between the pump and probe pulses. The fit, shown by the red line in Figure~\ref{fig:Intro}c, is done with the parameters $\omega_\textrm{L}=107\pm0.2$ rad ns$^{-1}$ and $T_2^*=170$\,ps.
The Fast Fourier Transform (FFT) spectrum of the KR dynamics is shown in the inset of Figure~\ref{fig:Intro}c. It has only one strong maximum at $\omega_\textrm{L}=109.3\pm8$ rad ns$^{-1}$, confirming that the coherent spin dynamics is dominated by one type of carriers. 

\begin{figure*}[!ht]
  \includegraphics[width=\textwidth]{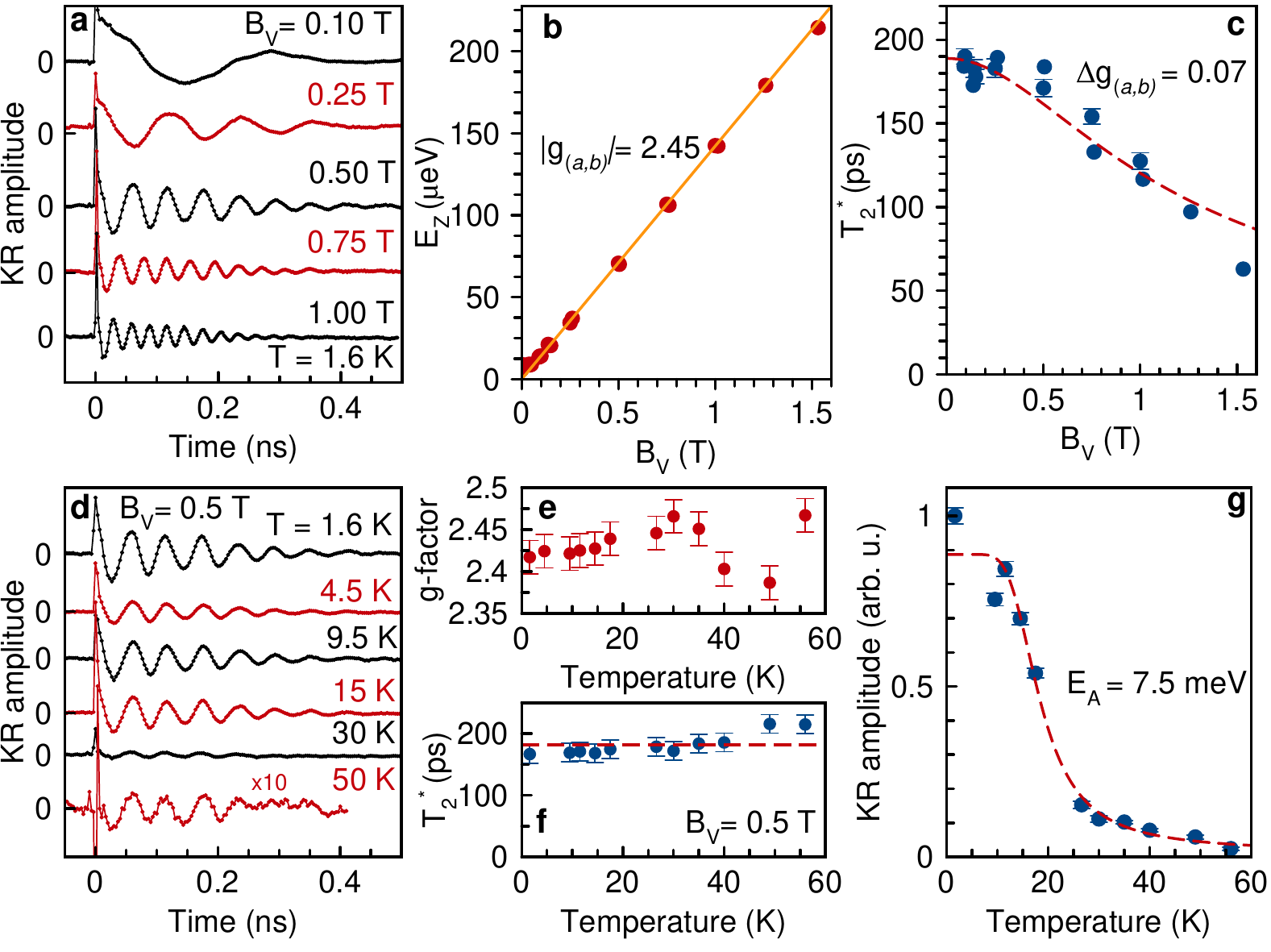}
  \caption{Coherent spin dynamics of electrons in (PEA)$_2$PbI$_4$.
(a) Time-resolved KR dynamics measured in various magnetic fields applied in Voigt geometry. $E_{\rm L}=2.3458$\,eV, $P=5$\,mW, and $T=1.6$\,K.  
(b) Zeeman splitting vs magnetic field, calculated from the Larmor frequency extracted from the signals in panel (a), (dots) together with a linear fit (line). $g_{(a,b)}$ denotes the in plane $g$-factor. 
(c) Spin dephasing time $T_2^*$ as function of the magnetic field (dots). Dashed line is a fit with Eq.~(\ref{eq:Dephasing_Time}) giving the fit parameters $T_{2,0}^*=190$\,ps and $\Delta g_{(a,b)}=0.06$.
(d) Time-resolved KR dynamics measured at various temperatures at $B_{\rm V}=0.5$\,T. $E_{\rm L}=2.345$\,eV, $P=3$\,mW. 
(e) Temperature dependence of the $g$-factor. 
(f) Temperature dependence of the spin dephasing time $T_2^*$. The average value (dashed line) is $T_2^* = 180$\,ps. 
(g) Temperature dependence of the KR amplitude at zero time delay (dots) and the corresponding fit (dashed line) with Eq.~(\ref{eq:KRamplitude}). 
The parameters given in the panels (e-g) are results of fits to the signals shown in panel (d), using Eq.~(\ref{eq:KerrRot}).} 
  \label{fig:BDep}
\end{figure*}

With increasing magnetic field strength the Larmor spin precession and the spin dephasing both accelerate. A set of KR dynamics traces for magnetic fields ranging from 0.1 up to 1.0~T is shown in Figure~\ref{fig:BDep}a. Through fits with Eq.~(\ref{eq:KerrRot}), the Larmor precession frequencies and spin dephasing times were evaluated. Figure~\ref{fig:BDep}b shows the magnetic field dependence of the Zeeman splitting, $E_{\rm Z}=\hbar\omega_{\rm L}$, which can be well described by a linear function without a notable offset at zero magnetic field. The slope of this dependence gives the value of the in-plane, spanned by a \& b axes, $g$-factor $|g_{(a,b)}|=2.45$. We will show below that it can be safely attributed to the electron $g$-factor, and therefore its sign should be positive~\cite{kirstein2022}. 
 
The magnetic field dependence of the spin dephasing time $T_2^*$ is shown in Figure~\ref{fig:BDep}c, shortening with increasing magnetic field. This is a well known behavior for spin ensembles, having a $g$-factor dispersion $\Delta g$~\cite{yakovlevCh6}. The dephasing time reads 
\begin{equation}
\label{eq:Dephasing_Time}
\frac{1}{T_2^{*}(B)} = \sqrt{\left(\frac{1}{T_{2,0}^{*}}\right)^2 + \left(\frac{\Delta g_{(a,b)} \mu_\textrm{B} B}{\hbar}\right)^2},
\end{equation}
with $T_{2,0}^*$ being the extrapolated zero magnetic field saturation time~\cite{evers2021}. The fit gives $T_{2,0}^*=190$\,ps and $\Delta g_{(a,b)}=0.07$. The latter corresponds to a relative $g$-factor variation, $\Delta g_{(a,b)}/g_{(a,b)}$, of 2.9\%. See further details in the Supporting Information, S8.

Note that in bulk lead halide perovskite crystals, like MAPbI$_3$, FA$_{0.9}$Cs$_{0.1}$PbI$_{2.8}$Br$_{0.2}$, FAPbBr$_3$, CsPbBr$_3$, typically two spin coherent signals with different Larmor frequencies are superimposed, leading to the observation of spin beats~\cite{kirstein2022}. They were identified as the spin precession signals of resident electrons and holes, which are localized at different crystal sites. The amplitudes of these TRKR signals were comparable evidencing that the concentration of resident electrons and holes are also close to each other.  

We attribute the observed spin precession in 2D (PEA)$_2$PbI$_4$ perovskite to the resident electrons. The spin dephasing time $T_2^*>190$\,ps exceeds greatly the exciton lifetime $\tau_{\textrm{Ref,} 1}=20$\,ps, which allows us to disregard the exciton spin precession or the electron spin precession within an exciton. There are several strong arguments, which allow us to favor the resident electrons over the resident holes. Among them are: (i) the $g$-factor value ranging close to the values obtained for electrons in bulk structures, (ii) its decreasing trend in (PEA)$_2$PbBr$_4$ structures (see \textbf{SI}) with increasing band gap \cite{kirstein2022}, and (iii) the absence of dynamic nuclear polarization (DNP), while a pronounced DNP occurs for holes in lead halide perovskites~\cite{kirstein2021}. More details are presented in the discussion.

The evolution of the KR dynamics with increasing temperature is shown in Figure~\ref{fig:BDep}d. The signal amplitude decreases strongly and vanishes at temperatures exceeding 50~K. Interestingly, the spin dephasing time (Figure~\ref{fig:BDep}f)  is only weakly affected by the temperature. Note that the insensitivity of the $g$-factor (Figure~\ref{fig:BDep}e) shows the maintenance of the character of the excited state.

The KR amplitude decrease with rising temperature (Figure~\ref{fig:BDep}g) can be fitted with an Arrhenius-type of dependence,  
\begin{equation}
\label{eq:KRamplitude}
\frac{1}{S(T)} = \frac{1}{S(0)}+w \exp{\left(-\frac{E_A}{k_{\rm B} T}\right)},
\end{equation} 
with the KR amplitude $S(T)$ evaluated at time equals zero, the Kerr amplitude $S(0)$ at zero temperature corresponding to a frozen phonon environment, the parameter $w$ characterizing the phonon interaction, and the Boltzmann constant $k_{\rm B}$. From a corresponding fit we get $E_A=7.5$\,meV as activation energy. This activation energy can be associated to several factors. Among them, 7.5\,meV (60.5\,cm$^{-1}$) corresponds to phonon modes, which are associated with the Pb-I bond and the inorganic cage - organic molecule bond motion~\cite{dhanabalan2020,do2020}. To remind, the valence and conduction bands in (PEA)$_2$PbI$_4$ are mainly formed by the antibonding orbital of Pb 6s - I 5p and Pb 6p, respectively. 

As additional aspect, the decrease of the electron spin signal is quite similar to that observed in perovskite lead halide bulk crystals~\cite{kirstein2021,jacoby2022}, for which carrier delocalization plays a major role in the temperature dependence.

\subsection{Longitudinal spin relaxation measured by spin inertia}

\begin{figure*}[!ht]
  \includegraphics[width=\textwidth]{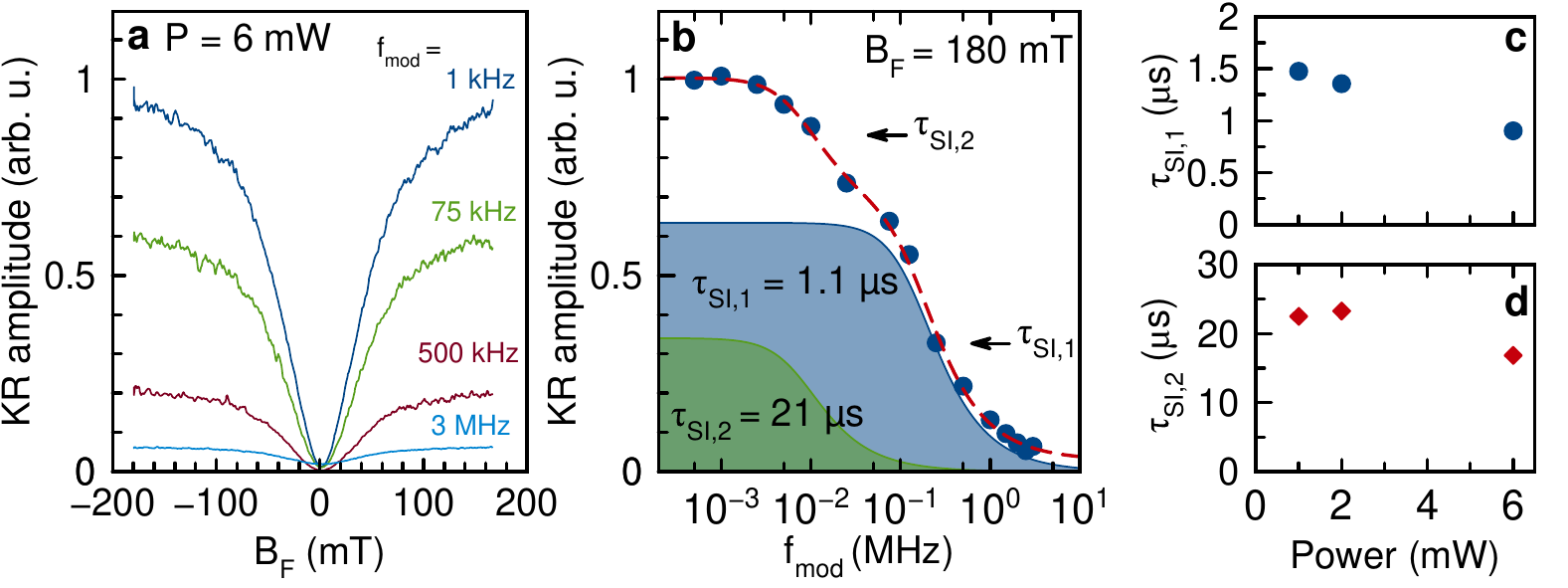}
\caption{(a) Set of polarization recovery curves (PRC) for various frequencies of the pump beam helicity modulation $f_\textrm{mod}$. The magnetic field is applied in the Faraday geometry. $E_{\rm L}=2.3594$\,eV, $P=6$\,mW, and $T=1.6$\,K. Note that the curves are not shifted vertically. 
  (b) Dependence of the KR amplitude at $B_{\rm F}=180$~mT on $f_\textrm{mod}$ (dots). The red dashed line is a fit with Eq.~(\ref{eq:inertia}), containing two longitudinal spin relaxation times. The individual contributions are shown by the colored areas. 
  (c,d) Power dependence of the spin relaxation times $\tau_\textrm{SI,1}$ and $\tau_\textrm{SI,2}$.}
  \label{fig:SpinIn}
\end{figure*}

In order to get information on the longitudinal spin relaxation dynamics, we perform experiments in a magnetic field applied in the Faraday geometry. No spin precession is expected in the longitudinal magnetic field ($B_\textrm{F}$) so that a clear separation between the longitudinal spin relaxation time, $T_1$, and the spin dephasing time, $T_2^*$,  can be made. We detect the spin polarization at a negative time delay of $-10$\,ps to get rid of any shorter spin dynamics that decay within the 13.2~ns after each laser pulse excitation. At zero magnetic field the TRKR signal is close to zero at this negative delay time, but it increases with $B_{\rm F}$ rising to 180~mT, Figure~\ref{fig:SpinIn}a. The field dependence has a characteristic Lorentzian shape, known as polarization recovery curve (PRC)~\cite{heisterkamp2015,smirnov2020}. Results for different modulation frequencies of the laser helicity, $f_\textrm{mod}$, are shown. At the frequency of 1\,kHz, the PRC curve has the full width at half maximum of 100\,mT. This width is nearly independent of $f_\textrm{mod}$. The increase of the spin polarization with rising magnetic field clearly indicates an increase of the spin lifetime with magnetic field which then exceeds $T_\textrm{R}=13.2$~ns. For $S(t=-10\, \textrm{ps}) = S_0 \exp(-t/T_1(B_\textrm{F}))$. One can see from this equation that  for a constant spin pumping $S_0$ to observe the increased PRC amplitude $T_1$ need to rise with the magnetic field. 

The faster modulation of the laser helicity, i.e. the increase of $f_\textrm{mod}$ up to 3~MHz, results in a considerable decrease of the KR amplitude (Figure~\ref{fig:SpinIn}a). This shows that the modulation period has become shorter than the spin relaxation time.  The results give access to the longitudinal spin relaxation time $T_1$ through the spin inertia method. Details of the spin inertia method are given in the Supporting Information, S2C. For that, we plot in Figure~\ref{fig:SpinIn}b the KR amplitude measured at $B_{\rm F}=180$~mT as a function of $f_\textrm{mod}$. The character of its decrease with increasing frequency evidences that two relaxation times are involved. We fit this dependence with 
\begin{equation}
\label{eq:inertia}
A_{\rm KR} = \sum_{n=\{1,2\}}\frac{S_{n}}{\sqrt{1+(2\pi f_\textrm{mod} \tau_{\textrm{SI},n})^2}},
\end{equation}
with the amplitude $S_n$ and the longitudinal spin relaxation times $\tau_{\textrm{SI},1}$ and $\tau_{\textrm{SI},2}$~\cite{belykh2019}. For the pump power of 6\,mW the fit yields $\tau_{\textrm{SI},1}=1.1\,\mu$s and $\tau_{\textrm{SI},2}=21\,\mu$s. 

By measuring the pump power dependencies of $\tau_{\textrm{SI},1}$ and $\tau_{\textrm{SI},2}$ times, we  extract the $T_1$ times in the limit of small excitation powers by using the equation $\tau_{\textrm{SI},n}^{-1} = T_{1,n}^{-1} + G_n P$, with the spin generation rate $G$ and the pump power $P$. Thereby, $T_{1,1}=1.7\,\mu$s and $T_{1,2}=25\,\mu$s are extracted, see Figures~\ref{fig:SpinIn}c,d. The observation of two distinct spin relaxation times evidences that the ensemble of resident electrons is not uniform and may be contributed by electrons subject to different localization conditions.

\subsection{$g$-factor anisotropy}

In bulk lead halide perovskites, the crystal symmetry can be reduced from cubic to tetragonal or orthorhombic due to a structural phase transition at low temperatures. This results in an anisotropy of the carrier $g$-factors~\cite{kirstein2022}. In 2D materials a symmetry reduction is caused by the spatial confinement, as the crystal axis perpendicular to the structural plane differs from the in-plane axes. Hence, analogously to the bulk case with low symmetry an anisotropic $g$-factor is expected in the present 2D structure. The $g$-factor is then described by $g_{i}$, where the $i = a,\,b,\,c$  index the respective crystallographic axes.  

In the studied sample (PEA)$_2$PbI$_4$, the monolayers are well oriented parallel to the substrate, i.e. the $c$-axis is oriented perpendicular to them, as confirmed by the XRD analysis given in the Supporting Information, S3. However, the sample shows domains of about 200\,nm size, see the Supporting Information, S4 and S5. Thus, the in-plane orientation may vary from domain to domain. Within the probed laser spot of 200\,$\mu$m the $a$- and $b$-axis orientations are assumed to be averaged out, therefore, we label them $g_{(a,b)}$.

\begin{figure}[!t]
  \includegraphics[width=\columnwidth]{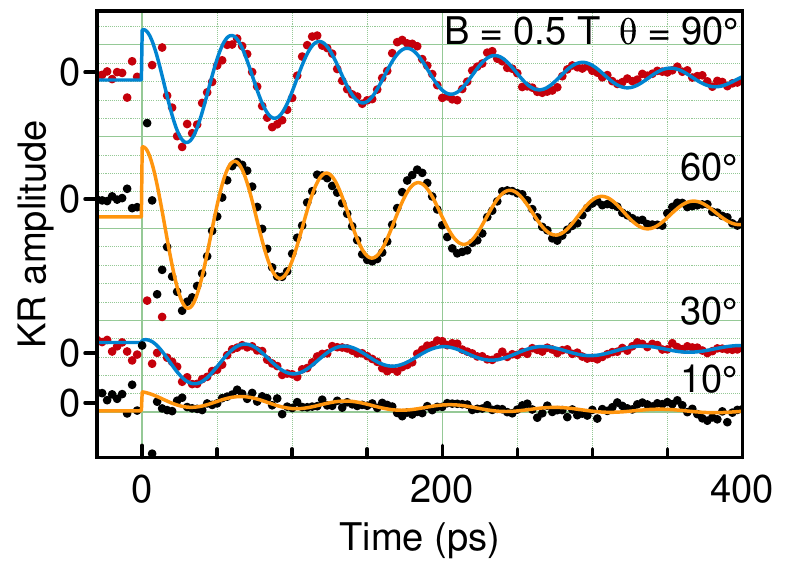}
  \caption{Time-resolved KR dynamics measured in a magnetic field of $B=0.5$\,T applied at different angles $\theta$. $\theta=90^\circ$ corresponds to the Voigt geometry ($\textbf{B}_{\rm V}\perp \textbf{k}$), while $\theta=0^\circ$ corresponds to the Faraday geometry ($\textbf{B}_{\rm F}\parallel \textbf{k}$), see Figure~\ref{fig:gfac}(a). Lines show fits to the experimental data with Eq.~(\ref{eq:KerrRot}). $E_{\rm L}=2.3536$\,eV, $P=1.5$\,mW, and $T=1.6$\,K.  }
  \label{fig:TRaniso}
\end{figure}

\begin{figure*}[!ht]
  \includegraphics[width=\textwidth]{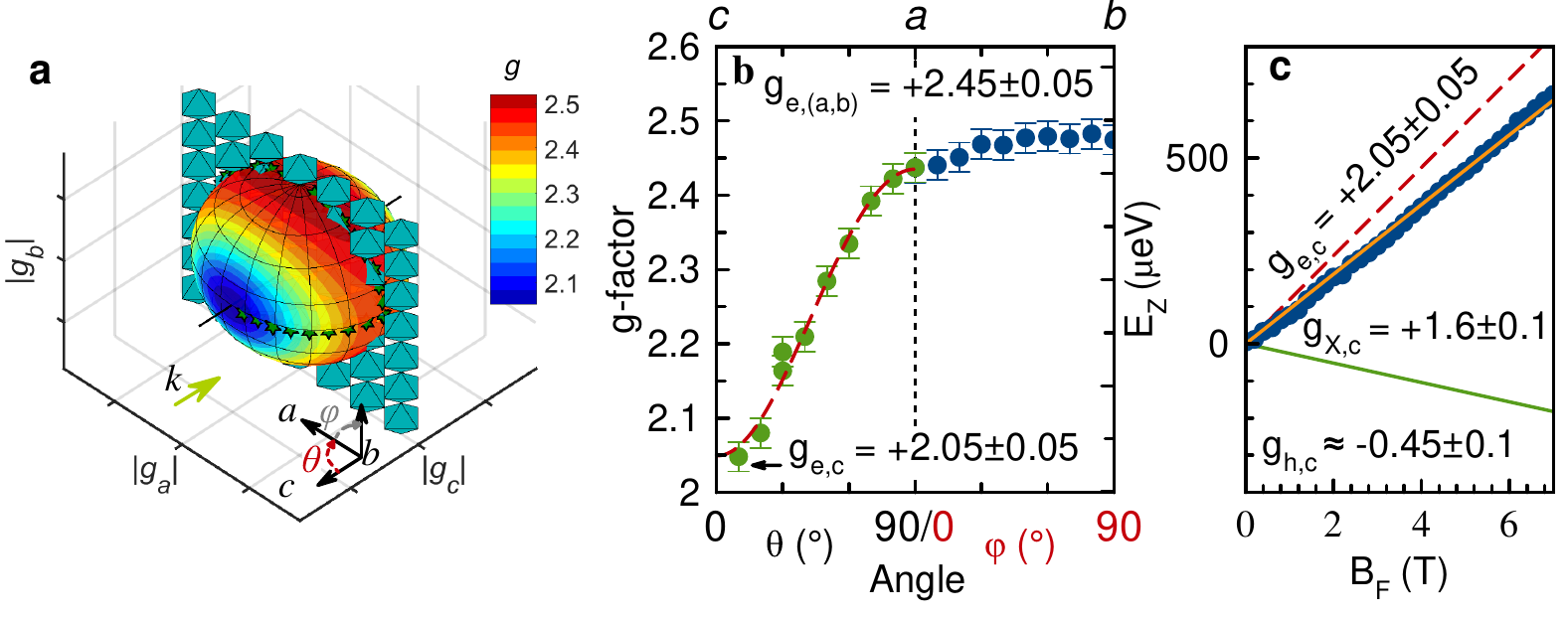}
  \caption{Anisotropy of the electron $g$-factor in (PEA)$_2$PbI$_4$. (a) Visualization of the $g$-factor tensor. The color code gives the effective $g$-factor value $g = \sqrt{g_a^2+g_b^2+g_c^2}$. A single plane of the perovskite corner shared octahedra is sketched as reference of the sample orientation. The axis cross gives the crystal planes and $k$ denotes the light propagation vector in the laboratory frame.  
(b) Angular dependence of the electron $g$-factor for the magnetic field rotation from the Faraday to the Voigt geometry (green dots) and for the in-plane rotation in the Voigt geometry (blue dots). Red line is a fit, $g(\theta) = \sqrt{(g_{\textrm{e},(a,b)}\sin{\theta})^2 + (g_{\textrm{e},c}\cos{\theta})^2}$, with $g$-factor values $g_{{\rm e},c}=+2.05$ and $g_{{\rm e},(a,b)}=+2.45$.  
(c) Zeeman splitting of the bright exciton as function of the applied magnetic field (dots), evaluated from the reflectivity spectra presented in the Supporting Information, S9. Orange line is a linear fit to the dependence with the exciton $g$-factor $g_{{\rm X},c}=+1.6\pm0.1$. Red dashed line gives the electron Zeeman splitting with the measured $g$-factor for the Faraday geometry ($c$-axis). Green line gives the hole Zeeman splitting calculated with $g_{{\rm h},c} = g_{{\rm X},c}-g_{{\rm e},c}$.}
  \label{fig:gfac}
\end{figure*}

For measurements of the $g$-factor anisotropy, the magnetic field direction was rotated, while the sample and light orientations were kept fixed. In Figure~\ref{fig:TRaniso} a set of KR dynamics is shown for rotation of the magnetic field from the Voigt geometry ($\theta=90^\circ$) towards the Faraday geometry ($\theta=0^\circ$). When approaching the Faraday geometry, the spin precession amplitude successively decreases, and shows, as long as resolvable, despite of its weakness a clear exponential decay. Also measurements for an in-plane magnetic field rotation in the Voigt geometry were performed. The measured electron $g$-factors for different orientation angles are summarized in Figure~\ref{fig:gfac}b. A pronounced $g$-factor anisotropy with $g_{{\rm e},c} = +2.05\pm0.05$ rising to $g_{{\rm e},(a,b)} = +2.45 \pm 0.05 $ is found for tilting the field direction from the out-of-plane towards the in-plane. The in-plane $g$-factor is almost constant, in line with the assumed domain averaging. The $g$-factor anisotropy is visualized as three-dimensional contour plot in Figure~\ref{fig:gfac}a. 

It is instructive to compare the electron and exciton $g$-factors. We measured the Zeeman splitting of the bright exciton, observed as splitting of its resonance dip in reflectivity spectra in a magnetic field applied in the Faraday geometry, see the Supporting Information, S9. Its field dependence is plotted in Figure~\ref{fig:gfac}c, and its slope gives $g_{{\rm X},c}=+1.6\pm0.1$. It is worth to note that in this experiment also the sign of the exciton $g$-factor can be identified, and it is positive. In the lead halide perovskites, for the bright exciton $g_{\rm X}=g_{\rm e}+g_{\rm h}$~\cite{kirstein2022}, which allows us to estimate $g_{{\rm h},c}=-0.45\pm0.1$. Note, that in literature values of $g_{{\rm X},c}=1.2\pm0.1$~\cite{posmyk2022,dyksik2020} and $g_{{\rm X},(a,b)}=1.9\pm0.5$~\cite{dyksik2021} were reported for the bright exciton in (PEA)$_2$PbI$_4$ monolayers ($n=1$). For convenience, we have collected the electron spin parameters measured for our sample in Table~\ref{tab:sum}.

\begin{table}%
\caption{Electron spin parameters measured for 2D (PEA)$_2$PbI$_4$ perovskites.}
\begin{tabular}{l||c|}
    & electron \\ \hline \hline 
$g_{\textrm{e},c}$ & $+2.05\pm0.05$ \\
$g_{\textrm{e},{(a,b)}}$ & $+2.45\pm0.05$ \\
$\Delta g_{{(a,b)}}$ &$0.06$   \\
$T_1$ & $1.7-25$\,$\mu$s    \\
$T_2^*$~~ & 190\,ps   
\end{tabular}
\label{tab:sum}
\end{table}

\subsection{Discussion}

Let us discuss the origin of the single frequency Larmor precession found in 2D (PEA)$_2$PbI$_4$. As we have argued above, the spin dephasing time is considerably longer than the exciton lifetime, which allows us to exclude the possibility of carrier spin precession within an exciton. The exciton itself can be also excluded, as its $g$-factor, which we measured directly, is considerably smaller than the value that we get from the TRKR experiments. Therefore, the spin coherent signal has to be provided by resident carriers, either the electrons or the holes.    

The first approach to distinguish electrons from holes is to use their different hyperfine interaction with the nuclear spins. In the lead halide perovskites, the hole spins interact about five to ten times more strongly with the nuclear spins than the electron spins~\cite{kirstein2021}. As a result, a considerable shift of the Larmor precession frequency of the holes can be achieved when the nuclear spins become dynamically polarized by the spin-oriented holes. We have performed such experiments for the studied (PEA)$_2$PbI$_4$ sample and did not found a measurable dynamic nuclear polarization. This is a strong argument in favor of the electron origin of spin precession. 

The second approach is to refer to the universal dependence of the electron and hole $g$-factors on the band gap energy, found recently experimentally and theoretically for bulk lead halide perovskites~\cite{kirstein2022}. We do not expect that this dependence will exactly match the results for 2D perovskites, as the electronic bands and the spin-orbit splitting are modified~\cite{quarti2020}. Still the main trends should hold. In bulk, for the band gap exceeding 1.8~eV both the electron and the hole $g$-factors are positive, but the electrons show a decrease with growing gap, while the holes show an increase. We performed TRKR experiments on a 2D (PEA)$_2$PbBr$_4$ reference sample ($n=1$), which has a larger band gap than the (PEA)$_2$PbI$_4$ sample. In this sample the exciton resonance has its energy at 3.050\,eV for $T=1.6$\,K. Again in TRKR only one precession frequency is found, here corresponding a smaller in-plane $g$-factor of 2.11, see the Supporting Information, S10. Compared to the $g$-factor $g_{{\rm e},(a,b)}=+2.45$ in the (PEA)$_2$PbI$_4$ sample, this supports the identification as electron spin as source of the precession.

The measured values of the electron $g$-factors in 2D perovskites are somewhat larger than the one predicted by the universal dependence for bulk crystals. Detailed model calculation are required here to clarify the details of the underlying band modifications. Note that also the dielectric environment can play a considerable role here via the chosen $A$ cation. The influence of the $A$ cation can be seen, for instance, from the finding that the exciton $g$-factor differs significantly within the class of the organics-based lead iodine monolayer perovskites: for (BA)$_2$PbI$_4$ it is $g_{{\rm X},c}=0.74+0.08$ and for (PEA)$_2$PbI$_4$, $g_{{\rm X},c}=1.2\pm0.1$ , with the respective 1s exciton energies of 2.52\,eV and 2.35\,eV, both obtained in high magnetic field studies~\cite{blancon2018,dyksik2020}. Note that in the present study we measured for (PEA)$_2$PbI$_4$ $g_{{\rm X}, c}=+1.6\pm0.1$.
 
In conclusion, the spin properties and spin dynamics of charge carriers and excitons have been studied for films composed of the two-dimensional perovskite (PEA)$_2$PbI$_4$ with $n=1$. The coherent spin dynamics measured at low temperatures and in magnetic fields are contributed exclusively by the electron Larmor spin precession. A significant anisotropy of the electron $g$-factor with the in-plane value of $+2.45$ and the out-of-plane value of $+2.05$ is found. By comparison with model calculations, these values will allow one to refine the values of the band parameters for the 2D perovskites. The electron spin dephasing time is 190~ps, while the longitudinal spin relaxation time in weak magnetic fields is as long as 25~$\mu$s. We envision an extension of these studies to 2D perovskites with a larger thickness of the perovskite layers, i.e. with $n=2,3,...$, and various organic barriers. In that way one could study the role of the quantum confinement and enhanced Coulomb interaction on the spin parameters of layered Ruddlesen-Popper perovskites and highlight their potential for spintronics applications.

\textbf{ASSOCIATED CONTENT}

The Supporting Information is available available free of charge at ... .

A detailed description of the methods, including the experimental apparatus, and the sample synthesis are provided. Further, the sample structure is analyzed via XRD, profilometry, and AFM. Additional experimental data such as the Kerr spectral dependence, TRKR pump power dependence, magneto reflectivity as well as data for a (PEA)$_2$PbBr$_4$ sample are presented (\textbf{PDF}). \\


\textbf{AUTHOR INFORMATION}

Corresponding Authors:

Email: erik.kirstein@tu-dortmund.de\\ 
Email: dmitri.yakovlev@tu-dortmund.de\\


\textbf{Notes}

The authors declare no competing interests.\\

\textbf{Acknowledgements}\\
We acknowledge the financial support by the Deutsche Forschungsgemeinschaft in the frame of the Priority Programme SPP 2196 (Project YA 65/26-1) and the International Collaboration Research Center TRR160 (project A1). The work at ETH Z\"urich (O.H. and M.V.K.) was financially supported by the Swiss National Science Foundation (grant agreement 186406, funded in conjunction with SPP2196 through DFG-SNSF bilateral program) and by the ETH Z\"urich through the ETH+ Project SynMatLab. We are thankful to Y. Shynkarenko, R. John and S. B\"ohme for providing us results of the profilometry and AFM measurements. We acknowledge M. A. Akmaev for assistance in the reflectivity measurements.



\begin{mcitethebibliography}{58}
\providecommand*\natexlab[1]{#1}
\providecommand*\mciteSetBstSublistMode[1]{}
\providecommand*\mciteSetBstMaxWidthForm[2]{}
\providecommand*\mciteBstWouldAddEndPuncttrue
  {\def\EndOfBibitem{\unskip.}}
\providecommand*\mciteBstWouldAddEndPunctfalse
  {\let\EndOfBibitem\relax}
\providecommand*\mciteSetBstMidEndSepPunct[3]{}
\providecommand*\mciteSetBstSublistLabelBeginEnd[3]{}
\providecommand*\EndOfBibitem{}
\mciteSetBstSublistMode{f}
\mciteSetBstMaxWidthForm{subitem}{(\alph{mcitesubitemcount})}
\mciteSetBstSublistLabelBeginEnd
  {\mcitemaxwidthsubitemform\space}
  {\relax}
  {\relax}

\bibitem[Ishihara \latin{et~al.}(1989)Ishihara, Takahashi, and
  Goto]{ishihara1989}
Ishihara,~T.; Takahashi,~J.; Goto,~T. Exciton state in two-dimensional
  perovskite semiconductor (C$_{10}$H$_{21}$NH$_3$)$_2$PbI$_4$. \textit{Solid State Commun.}
  \textbf{1989}, \textit{69}, 933--936, DOI:
  \doi{10.1016/0038-1098(89)90935-6}\relax
\mciteBstWouldAddEndPuncttrue
\mciteSetBstMidEndSepPunct{\mcitedefaultmidpunct}
{\mcitedefaultendpunct}{\mcitedefaultseppunct}\relax
\EndOfBibitem
\bibitem[Ishihara \latin{et~al.}(1990)Ishihara, Takahashi, and
  Goto]{ishihara1990}
Ishihara,~T.; Takahashi,~J.; Goto,~T. Optical properties due to electronic
  transitions in two-dimensional semiconductors
  ({C}$_n$H$_{(2n+1)}${NH$_3$}){PbI$_4$}. \textit{Phys. Rev. B} \textbf{1990},
  \textit{42}, 11099--11107, DOI: \doi{10.1103/PhysRevB.42.11099}\relax
\mciteBstWouldAddEndPuncttrue
\mciteSetBstMidEndSepPunct{\mcitedefaultmidpunct}
{\mcitedefaultendpunct}{\mcitedefaultseppunct}\relax
\EndOfBibitem
\bibitem[Kataoka \latin{et~al.}(1993)Kataoka, Kondo, Ito, Sasaki, Uchida, and
  Miura]{kataoka1993}
Kataoka,~T.; Kondo,~T.; Ito,~R.; Sasaki,~S.; Uchida,~K.; Miura,~N.
  Magneto-optical study on excitonic spectra in
  ({C}$_6$H$_{13}$NH$_3$)$_2$PbI$_4$. \textit{Phys. Rev. B} \textbf{1993},
  \textit{47}, 2010--2018, DOI: \doi{10.1103/PhysRevB.47.2010}\relax
\mciteBstWouldAddEndPuncttrue
\mciteSetBstMidEndSepPunct{\mcitedefaultmidpunct}
{\mcitedefaultendpunct}{\mcitedefaultseppunct}\relax
\EndOfBibitem
\bibitem[Muljarov \latin{et~al.}(1995)Muljarov, Tikhodeev, Gippius, and
  Ishihara]{muljarov1995}
Muljarov,~E.~A.; Tikhodeev,~S.~G.; Gippius,~N.~A.; Ishihara,~T. Excitons in
  self-organized semiconductor/insulator superlattices: {PbI}-based perovskite
  compounds. \textit{Phys. Rev. B} \textbf{1995}, \textit{51}, 14370--14378, DOI:
  \doi{10.1103/PhysRevB.51.14370}\relax
\mciteBstWouldAddEndPuncttrue
\mciteSetBstMidEndSepPunct{\mcitedefaultmidpunct}
{\mcitedefaultendpunct}{\mcitedefaultseppunct}\relax
\EndOfBibitem
\bibitem[Blancon \latin{et~al.}(2020)Blancon, Even, Stoumpos, Kanatzidis, and
  Mohite]{blancon2020}
Blancon,~J.-C.; Even,~J.; Stoumpos,~C.~C.; Kanatzidis,~M.~G.; Mohite,~A.~D.
  Semiconductor physics of organic–inorganic {2D} halide perovskites.
  \textit{Nat. Nanotechnol.} \textbf{2020}, \textit{15}, 969--985, DOI:
  \doi{10.1038/s41565-020-00811-1}\relax
\mciteBstWouldAddEndPuncttrue
\mciteSetBstMidEndSepPunct{\mcitedefaultmidpunct}
{\mcitedefaultendpunct}{\mcitedefaultseppunct}\relax
\EndOfBibitem
\bibitem[Wang \latin{et~al.}(2018)Wang, Chernikov, Glazov, Heinz, Marie, Amand,
  and Urbaszek]{wang2018}
Wang,~G.; Chernikov,~A.; Glazov,~M.~M.; Heinz,~T.~F.; Marie,~X.; Amand,~T.;
  Urbaszek,~B. Colloquium: {Excitons} in atomically thin transition metal
  dichalcogenides. \textit{RMP} \textbf{2018}, \textit{90}, 021001, DOI:
  \doi{10.1103/RevModPhys.90.021001}\relax
\mciteBstWouldAddEndPuncttrue
\mciteSetBstMidEndSepPunct{\mcitedefaultmidpunct}
{\mcitedefaultendpunct}{\mcitedefaultseppunct}\relax
\EndOfBibitem
\bibitem[Mao \latin{et~al.}(2018)Mao, Stoumpos, and Kanatzidis]{mao2018}
Mao,~L.; Stoumpos,~C.~C.; Kanatzidis,~M.~G. Two-{dimensional} {hybrid} {halide}
  {perovskites}: {principles} and {promises}. \textit{J. Am. Chem. Soc.}
  \textbf{2018}, \textit{141}, DOI: \doi{10.1021/jacs.8b10851}\relax
\mciteBstWouldAddEndPuncttrue
\mciteSetBstMidEndSepPunct{\mcitedefaultmidpunct}
{\mcitedefaultendpunct}{\mcitedefaultseppunct}\relax
\EndOfBibitem
\bibitem[\L{}empicka-Mirek \latin{et~al.}(2022)\L{}empicka-Mirek, Kr\'ol,
  Sigurdsson, Wincukiewicz, Morawiak, Mazur, Muszyński, Piecek, Kula,
  Stefaniuk, Kami\'nska, De~Marco, Lagoudakis, Ballarini, Sanvitto, Szczytko,
  and Piętka]{lempicka-mirek2022}
\L{}empicka-Mirek,~K.;  Krol, M.;  Sigurdsson, H.;  Wincukiewicz, A.;  Morawiak, P.;  Mazur, R.;  Muszynski, M.;  Piecek, W.;  Kula, P.;  Stefaniuk, T.;  Kaminska, M.;  De Marco, L.;  Lagoudakis, P.G.;  Ballarini, D.;  Sanvitto, D.; Szczytko, J.; Pietka, B. 
Electrically tunable {Berry} curvature
  and strong light-matter coupling in birefringent perovskite microcavities at
  room temperature. \textit{arXiv [cond-mat, physics:physics]} \textbf{2022},
  DOI: \doi{10.48550/arXiv.2203.05289}\relax
\mciteBstWouldAddEndPuncttrue
\mciteSetBstMidEndSepPunct{\mcitedefaultmidpunct}
{\mcitedefaultendpunct}{\mcitedefaultseppunct}\relax
\EndOfBibitem
\bibitem[Hoye \latin{et~al.}(2022)Hoye, Hidalgo, Jagt, Correa-Baena, Fix, and
  MacManus-Driscoll]{hoye2022}
Hoye,~R. L.~Z.; Hidalgo,~J.; Jagt,~R.~A.; Correa-Baena,~J.-P.; Fix,~T.;
  MacManus-Driscoll,~J.~L. The {role} of {dimensionality} on the
  {optoelectronic} {properties} of {oxide} and {halide} {perovskites}, and
  their {halide} {derivatives}. \textit{Adv. Energy Mater.} \textbf{2022},
  \textit{12}, 2100499, DOI: \doi{10.1002/aenm.202100499}\relax
\mciteBstWouldAddEndPuncttrue
\mciteSetBstMidEndSepPunct{\mcitedefaultmidpunct}
{\mcitedefaultendpunct}{\mcitedefaultseppunct}\relax
\EndOfBibitem
\bibitem[Vardeny \latin{et~al.}(2022)Vardeny, Beard, Vardeny, and
  Beard]{vardeny2022c}
Vardeny,~Z.~V.; Beard,~M.~C.; Vardeny,~Z.~V.; Beard,~M.~C. \textit{Hybrid
  {Organic} {Inorganic} {Perovskites}: {Physical} {Properties} and
  {ApplicationsVolume} 3: {Spin} {Response} of {Hybrid} {Organic} {Inorganic}
  {Perovskites}}; World Scientific, 2022; Vol.~18; DOI:
  \doi{10.1142/12387-vol3}\relax
\mciteBstWouldAddEndPuncttrue
\mciteSetBstMidEndSepPunct{\mcitedefaultmidpunct}
{\mcitedefaultendpunct}{\mcitedefaultseppunct}\relax
\EndOfBibitem
\bibitem[Vardeny \latin{et~al.}(2022)Vardeny, Beard, Vardeny, and
  Beard]{vardeny2022d}
Vardeny,~Z.~V.; Beard,~M.~C.; Vardeny,~Z.~V.; Beard,~M.~C. \textit{Hybrid
  {Organic} {Inorganic} {Perovskites}: {Physical} {Properties} and
  {ApplicationsVolume} 4: {Hybrid} {Organic} {Inorganic} {Perovskite}
  {Applications}}; World Scientific, 2022; Vol.~18; DOI:
  \doi{10.1142/12387-vol4}\relax
\mciteBstWouldAddEndPuncttrue
\mciteSetBstMidEndSepPunct{\mcitedefaultmidpunct}
{\mcitedefaultendpunct}{\mcitedefaultseppunct}\relax
\EndOfBibitem
\bibitem[Ricciardulli \latin{et~al.}(2021)Ricciardulli, Yang, Smet, and
  Saliba]{ricciardulli2021}
Ricciardulli,~A.~G.; Yang,~S.; Smet,~J.~H.; Saliba,~M. Emerging perovskite
  monolayers. \textit{Nat. Mater.} \textbf{2021}, \textit{20}, 1325--1336, DOI:
  \doi{10.1038/s41563-021-01029-9}\relax
\mciteBstWouldAddEndPuncttrue
\mciteSetBstMidEndSepPunct{\mcitedefaultmidpunct}
{\mcitedefaultendpunct}{\mcitedefaultseppunct}\relax
\EndOfBibitem
\bibitem[Sidhik \latin{et~al.}(2021)Sidhik, Wang, Li, Zhang, Zhong, Agrawal,
  Hadar, Spanopoulos, Mishra, Traor\'e, Samani, Katan, Marciel, Blancon, Even,
  Kahn, Kanatzidis, and Mohite]{sidhik2021}
Sidhik,~S. \latin{et~al.}  High-phase purity two-dimensional perovskites with
  17.3\% efficiency enabled by interface engineering of hole transport layer.
  \textit{Cell Rep. Phys. Sci.} \textbf{2021}, \textit{2}, 100601, DOI:
  \doi{10.1016/j.xcrp.2021.100601}\relax
\mciteBstWouldAddEndPuncttrue
\mciteSetBstMidEndSepPunct{\mcitedefaultmidpunct}
{\mcitedefaultendpunct}{\mcitedefaultseppunct}\relax
\EndOfBibitem
\bibitem[Seitz \latin{et~al.}(2020)Seitz, Magdaleno, Alc\'azar-Cano,
  Mel\'endez, Lubbers, Walraven, Pakdel, Prada, Delgado-Buscalioni, and
  Prins]{seitz2020}
Seitz,~M.; Magdaleno,~A.~J.; Alc\'azar-Cano,~N.; Mel\'endez,~M.;
  Lubbers,~T.~J.; Walraven,~S.~W.; Pakdel,~S.; Prada,~E.;
  Delgado-Buscalioni,~R.; Prins,~F. Exciton diffusion in two-dimensional
  metal-halide perovskites. \textit{Nat. Commun.} \textbf{2020}, \textit{11}, 2035,
  DOI: \doi{10.1038/s41467-020-15882-w}\relax
\mciteBstWouldAddEndPuncttrue
\mciteSetBstMidEndSepPunct{\mcitedefaultmidpunct}
{\mcitedefaultendpunct}{\mcitedefaultseppunct}\relax
\EndOfBibitem
\bibitem[Grancini \latin{et~al.}(2017)Grancini, Roldán-Carmona, Zimmermann,
  Mosconi, Lee, Martineau, Narbey, Oswald, De~Angelis, Graetzel, and
  Nazeeruddin]{grancini2017}
Grancini,~G.; Roldán-Carmona,~C.; Zimmermann,~I.; Mosconi,~E.; Lee,~X.;
  Martineau,~D.; Narbey,~S.; Oswald,~F.; De~Angelis,~F.; Graetzel,~M.;
  Nazeeruddin,~M.~K. One-{year} stable perovskite solar cells by {2D}/{3D}
  interface engineering. \textit{Nat. Commun.} \textbf{2017}, \textit{8}, 15684,
  DOI: \doi{10.1038/ncomms15684}\relax
\mciteBstWouldAddEndPuncttrue
\mciteSetBstMidEndSepPunct{\mcitedefaultmidpunct}
{\mcitedefaultendpunct}{\mcitedefaultseppunct}\relax
\EndOfBibitem
\bibitem[Thouin \latin{et~al.}(2018)Thouin, Neutzner, Cortecchia, Dragomir,
  Soci, Salim, Lam, Leonelli, Petrozza, Kandada, and Silva]{thouin2018}
Thouin,~F.; Neutzner,~S.; Cortecchia,~D.; Dragomir,~V.~A.; Soci,~C.; Salim,~T.;
  Lam,~Y.~M.; Leonelli,~R.; Petrozza,~A.; Kandada,~A. R.~S.; Silva,~C. Stable
  biexcitons in two-dimensional metal-halide perovskites with strong dynamic
  lattice disorder. \textit{Phys. Rev. Mater.} \textbf{2018}, \textit{2}, 034001,
  DOI: \doi{10.1103/PhysRevMaterials.2.034001}\relax
\mciteBstWouldAddEndPuncttrue
\mciteSetBstMidEndSepPunct{\mcitedefaultmidpunct}
{\mcitedefaultendpunct}{\mcitedefaultseppunct}\relax
\EndOfBibitem
\bibitem[Qin \latin{et~al.}(2022)Qin, Li, Gao, Chen, Li, Xu, Li, Jiang, Li, Wu,
  Quan, Ye, Zhang, Lin, Pedesseau, Even, Lu, and Bu]{qin2022}
Qin,~Y. \latin{et~al.}  Dangling {octahedra} {enable} {edge} {states} in {2D}
  {lead} {halide} {perovskites}. \textit{Adv. Mater.} \textbf{2022}, \textit{34},
  2201666, DOI: \doi{10.1002/adma.202201666}\relax
\mciteBstWouldAddEndPuncttrue
\mciteSetBstMidEndSepPunct{\mcitedefaultmidpunct}
{\mcitedefaultendpunct}{\mcitedefaultseppunct}\relax
\EndOfBibitem
\bibitem[Blancon \latin{et~al.}(2018)Blancon, Stier, Tsai, Nie, Stoumpos,
  Traor\'e, Pedesseau, Kepenekian, Katsutani, Noe, Kono, Tretiak, Crooker,
  Katan, Kanatzidis, Crochet, Even, and Mohite]{blancon2018}
Blancon,~J.-C. \latin{et~al.}  Scaling law for excitons in {2D} perovskite
  quantum wells. \textit{Nat. Commun.} \textbf{2018}, \textit{9}, 2254, DOI:
  \doi{10.1038/s41467-018-04659-x}\relax
\mciteBstWouldAddEndPuncttrue
\mciteSetBstMidEndSepPunct{\mcitedefaultmidpunct}
{\mcitedefaultendpunct}{\mcitedefaultseppunct}\relax
\EndOfBibitem
\bibitem[Dyksik \latin{et~al.}(2020)Dyksik, Duim, Zhu, Yang, Gen, Kohama,
  Adjokatse, Maude, Loi, Egger, Baranowski, and Plochocka]{dyksik2020}
Dyksik,~M.; Duim,~H.; Zhu,~X.; Yang,~Z.; Gen,~M.; Kohama,~Y.; Adjokatse,~S.;
  Maude,~D.~K.; Loi,~M.~A.; Egger,~D.~A.; Baranowski,~M.; Plochocka,~P. Broad
  {tunability} of {carrier} {effective} {masses} in {two}-{dimensional}
  {halide} {perovskites}. \textit{ACS Energy Lett.} \textbf{2020}, \textit{5},
  3609--3616, DOI: \doi{10.1021/acsenergylett.0c01758}\relax
\mciteBstWouldAddEndPuncttrue
\mciteSetBstMidEndSepPunct{\mcitedefaultmidpunct}
{\mcitedefaultendpunct}{\mcitedefaultseppunct}\relax
\EndOfBibitem
\bibitem[Dyksik \latin{et~al.}(2021)Dyksik, Duim, Maude, Baranowski, Loi, and
  Plochocka]{dyksik2021}
Dyksik,~M.; Duim,~H.; Maude,~D.~K.; Baranowski,~M.; Loi,~M.~A.; Plochocka,~P.
  Brightening of dark excitons in {2D} perovskites. \textit{Sci. Adv.}
  \textbf{2021}, \textit{7}, eabk0904,  DOI: \doi{10.1126/sciadv.abk0904}\relax
\mciteBstWouldAddEndPuncttrue
\mciteSetBstMidEndSepPunct{\mcitedefaultmidpunct}
{\mcitedefaultendpunct}{\mcitedefaultseppunct}\relax
\EndOfBibitem
\bibitem[Do \latin{et~al.}(2020)Do, Granados~del \'Aguila, Zhang, Xing, Liu,
  Prosnikov, Gao, Chang, Christianen, and Xiong]{do2020b}
Do,~T. T.~H.; Granados~del \'Aguila,~A.; Zhang,~D.; Xing,~J.; Liu,~S.;
  Prosnikov,~M.~A.; Gao,~W.; Chang,~K.; Christianen,~P. C.~M.; Xiong,~Q. Bright
  {exciton} {fine}-{structure} in {two}-{dimensional} {lead} {halide}
  {perovskites}. \textit{Nano Lett.} \textbf{2020}, \textit{20}, 5141--5148, DOI:
  \doi{10.1021/acs.nanolett.0c01364}\relax
\mciteBstWouldAddEndPuncttrue
\mciteSetBstMidEndSepPunct{\mcitedefaultmidpunct}
{\mcitedefaultendpunct}{\mcitedefaultseppunct}\relax
\EndOfBibitem
\bibitem[Surrente \latin{et~al.}(2021)Surrente, Baranowski, and
  Plochocka]{surrente2021}
Surrente,~A.; Baranowski,~M.; Plochocka,~P. Perspective on the physics of
  two-dimensional perovskites in high magnetic field. \textit{Appl. Phys. Lett.}
  \textbf{2021}, \textit{118}, 170501, DOI: \doi{10.1063/5.0048490}\relax
\mciteBstWouldAddEndPuncttrue
\mciteSetBstMidEndSepPunct{\mcitedefaultmidpunct}
{\mcitedefaultendpunct}{\mcitedefaultseppunct}\relax
\EndOfBibitem
\bibitem[Giovanni \latin{et~al.}(2018)Giovanni, Chong, Liu, Dewi, Yin, Lekina,
  Shen, Mathews, Gan, and Sum]{giovanni2018}
Giovanni,~D.; Chong,~W.~K.; Liu,~Y. Y.~F.; Dewi,~H.~A.; Yin,~T.; Lekina,~Y.;
  Shen,~Z.~X.; Mathews,~N.; Gan,~C.~K.; Sum,~T.~C. Coherent {spin} and
  {quasiparticle} {dynamics} in {solution}-{processed} {layered} {2D} {lead}
  {halide} {perovskites}. \textit{Adv. Sci.} \textbf{2018}, \textit{5}, 1800664,
  DOI: \doi{10.1002/advs.201800664}\relax
\mciteBstWouldAddEndPuncttrue
\mciteSetBstMidEndSepPunct{\mcitedefaultmidpunct}
{\mcitedefaultendpunct}{\mcitedefaultseppunct}\relax
\EndOfBibitem
\bibitem[Chen \latin{et~al.}(2021)Chen, Lu, Wang, Zhai, Lunin, Sercel, and
  Beard]{chen2021}
Chen,~X.; Lu,~H.; Wang,~K.; Zhai,~Y.; Lunin,~V.; Sercel,~P.~C.; Beard,~M.~C.
  Tuning {spin}-{polarized} {lifetime} in {two}-{dimensional}
  {metal}–{halide} {perovskite} through {exciton} {binding} {energy}.
  \textit{J. ACS} \textbf{2021}, \textit{143}, 19438--19445, DOI:
  \doi{10.1021/jacs.1c08514}\relax
\mciteBstWouldAddEndPuncttrue
\mciteSetBstMidEndSepPunct{\mcitedefaultmidpunct}
{\mcitedefaultendpunct}{\mcitedefaultseppunct}\relax
\EndOfBibitem
\bibitem[Bourelle \latin{et~al.}(2022)Bourelle, Camargo, Ghosh, Neumann, van~de
  Goor, Shivanna, Winkler, Cerullo, and Deschler]{bourelle2022}
Bourelle,~S.~A.; Camargo,~F. V.~A.; Ghosh,~S.; Neumann,~T.; van~de Goor,~T.
  W.~J.; Shivanna,~R.; Winkler,~T.; Cerullo,~G.; Deschler,~F. Optical control
  of exciton spin dynamics in layered metal halide perovskites via polaronic
  state formation. \textit{Nat. Commun.} \textbf{2022}, \textit{13}, 3320, DOI:
  \doi{10.1038/s41467-022-30953-w}\relax
\mciteBstWouldAddEndPuncttrue
\mciteSetBstMidEndSepPunct{\mcitedefaultmidpunct}
{\mcitedefaultendpunct}{\mcitedefaultseppunct}\relax
\EndOfBibitem
\bibitem[Bourelle \latin{et~al.}(2020)Bourelle, Shivanna, Camargo, Ghosh,
  Gillett, Senanayak, Feldmann, Eyre, Ashoka, van~de Goor, Abolins, Winkler,
  Cerullo, Friend, and Deschler]{bourelle2020}
Bourelle,~S.~A.; Shivanna,~R.; Camargo,~F. V.~A.; Ghosh,~S.; Gillett,~A.~J.;
  Senanayak,~S.~P.; Feldmann,~S.; Eyre,~L.; Ashoka,~A.; van~de Goor,~T. W.~J.;
  Abolins,~H.; Winkler,~T.; Cerullo,~G.; Friend,~R.~H.; Deschler,~F. How
  {exciton} {interactions} {control} {spin}-{depolarization} in {layered}
  {hybrid} {perovskites}. \textit{Nano Lett.} \textbf{2020}, \textit{20},
  5678--5685, DOI: \doi{10.1021/acs.nanolett.0c00867}\relax
\mciteBstWouldAddEndPuncttrue
\mciteSetBstMidEndSepPunct{\mcitedefaultmidpunct}
{\mcitedefaultendpunct}{\mcitedefaultseppunct}\relax
\EndOfBibitem
\bibitem[Belykh \latin{et~al.}(2019)Belykh, Yakovlev, Glazov, Grigoryev,
  Hussain, Rautert, Dirin, Kovalenko, and Bayer]{belykh2019}
Belykh,~V.~V.; Yakovlev,~D.~R.; Glazov,~M.~M.; Grigoryev,~P.~S.; Hussain,~M.;
  Rautert,~J.; Dirin,~D.~N.; Kovalenko,~M.~V.; Bayer,~M. Coherent spin dynamics
  of electrons and holes in {CsPbBr}$_3$ perovskite crystals. \textit{Nat.
  Commun.} \textbf{2019}, \textit{10}, 673, DOI:
  \doi{10.1038/s41467-019-08625-z}\relax
\mciteBstWouldAddEndPuncttrue
\mciteSetBstMidEndSepPunct{\mcitedefaultmidpunct}
{\mcitedefaultendpunct}{\mcitedefaultseppunct}\relax
\EndOfBibitem
\bibitem[Kirstein \latin{et~al.}(2022)Kirstein, Yakovlev, Glazov, Evers,
  Zhukov, Belykh, Kopteva, Kudlacik, Nazarenko, Dirin, Kovalenko, and
  Bayer]{kirstein2021}
Kirstein,~E.; Yakovlev,~D.~R.; Glazov,~M.~M.; Evers,~E.; Zhukov,~E.~A.;
  Belykh,~V.~V.; Kopteva,~N.~E.; Kudlacik,~D.; Nazarenko,~O.; Dirin,~D.~N.;
  Kovalenko,~M.~V.; Bayer,~M. Lead-{dominated} {hyperfine} {interaction}
  {impacting} the {carrier} {spin} {dynamics} in {halide} {perovskites}.
  \textit{Adv. Mater.} \textbf{2022}, \textit{34}, 2105263, DOI:
  \doi{10.1002/adma.202105263}\relax
\mciteBstWouldAddEndPuncttrue
\mciteSetBstMidEndSepPunct{\mcitedefaultmidpunct}
{\mcitedefaultendpunct}{\mcitedefaultseppunct}\relax
\EndOfBibitem
\bibitem[Kirstein \latin{et~al.}(2022)Kirstein, Yakovlev, Glazov, Zhukov,
  Kudlacik, Kalitukha, Sapega, Dimitriev, Semina, Nestoklon, Ivchenko, Kopteva,
  Dirin, Nazarenko, Kovalenko, Baumann, Höcker, Dyakonov, and
  Bayer]{kirstein2022}
Kirstein,~E. \latin{et~al.}  The {Land\'e} factors of electrons and holes in
  lead halide perovskites: universal dependence on the band gap. \textit{Nat.
  Commun.} \textbf{2022}, \textit{13}, 3062, DOI:
  \doi{10.1038/s41467-022-30701-0}\relax
\mciteBstWouldAddEndPuncttrue
\mciteSetBstMidEndSepPunct{\mcitedefaultmidpunct}
{\mcitedefaultendpunct}{\mcitedefaultseppunct}\relax
\EndOfBibitem
\bibitem[Kirstein \latin{et~al.}(2022)Kirstein, Yakovlev, Zhukov, H\"ocker,
  Dyakonov, and Bayer]{kirstein2022b}
Kirstein,~E.; Yakovlev,~D.~R.; Zhukov,~E.~A.; H\"ocker,~J.; Dyakonov,~V.;
  Bayer,~M. Spin {dynamics} of {electrons} and {holes} {interacting} with
  {nuclei} in {MAPbI$_3$} {perovskite} {single} {crystals}. \textit{ACS
  Photonics} \textbf{2022}, \textit{9}, 1375--1384, DOI:
  \doi{10.1021/acsphotonics.2c00096}\relax
\mciteBstWouldAddEndPuncttrue
\mciteSetBstMidEndSepPunct{\mcitedefaultmidpunct}
{\mcitedefaultendpunct}{\mcitedefaultseppunct}\relax
\EndOfBibitem
\bibitem[Odenthal \latin{et~al.}(2017)Odenthal, Talmadge, Gundlach, Wang,
  Zhang, Sun, Yu, Valy~Vardeny, and Li]{odenthal2017}
Odenthal,~P.; Talmadge,~W.; Gundlach,~N.; Wang,~R.; Zhang,~C.; Sun,~D.;
  Yu,~Z.-G.; Valy~Vardeny,~Z.; Li,~Y.~S. Spin-polarized exciton quantum beating
  in hybrid organic–inorganic perovskites. \textit{Nature Physics}
  \textbf{2017}, \textit{13}, 894--899, DOI: \doi{10.1038/nphys4145}\relax
\mciteBstWouldAddEndPuncttrue
\mciteSetBstMidEndSepPunct{\mcitedefaultmidpunct}
{\mcitedefaultendpunct}{\mcitedefaultseppunct}\relax
\EndOfBibitem
\bibitem[Garcia-Arellano \latin{et~al.}(2021)Garcia-Arellano, Tripp\'e-Allard,
  Legrand, Barisien, Garrot, Deleporte, Bernardot, Testelin, and
  Chamarro]{garcia-arellano2021}
Garcia-Arellano,~G.; Tripp\'e-Allard,~G.; Legrand,~L.; Barisien,~T.;
  Garrot,~D.; Deleporte,~E.; Bernardot,~F.; Testelin,~C.; Chamarro,~M. Energy
  {tuning} of {electronic} {spin} {coherent} {evolution} in {methylammonium}
  {lead} {iodide} {perovskites}. \textit{J. Phys. Chem. Lett.} \textbf{2021}, \textit{12}, 8272--8279, DOI:
  \doi{10.1021/acs.jpclett.1c01790}\relax
\mciteBstWouldAddEndPuncttrue
\mciteSetBstMidEndSepPunct{\mcitedefaultmidpunct}
{\mcitedefaultendpunct}{\mcitedefaultseppunct}\relax
\EndOfBibitem
\bibitem[Garcia-Arellano \latin{et~al.}(2022)Garcia-Arellano, Tripp\'e-Allard,
  Campos, Bernardot, Legrand, Garrot, Deleporte, Testelin, and
  Chamarro]{garcia-arellano2022}
Garcia-Arellano,~G.; Tripp\'e-Allard,~G.; Campos,~T.; Bernardot,~F.;
  Legrand,~L.; Garrot,~D.; Deleporte,~E.; Testelin,~C.; Chamarro,~M. Unexpected
  {anisotropy} of the {electron} and {hole} {Land\'e} g-{factors} in
  {perovskite} {CH$_3$NH$_3$PbI$_3$} {polycrystalline} {films}. \textit{Nanomaterials}
  \textbf{2022}, \textit{12}, 1399, DOI: \doi{10.3390/nano12091399}\relax
\mciteBstWouldAddEndPuncttrue
\mciteSetBstMidEndSepPunct{\mcitedefaultmidpunct}
{\mcitedefaultendpunct}{\mcitedefaultseppunct}\relax
\EndOfBibitem
\bibitem[Crane \latin{et~al.}(2020)Crane, Jacoby, Cohen, Huang, Luscombe, and
  Gamelin]{crane2020}
Crane,~M.~J.; Jacoby,~L.~M.; Cohen,~T.~A.; Huang,~Y.; Luscombe,~C.~K.;
  Gamelin,~D.~R. Coherent {spin} {precession} and {lifetime}-{limited} {spin}
  {dephasing} in {CsPbBr$_3$} {perovskite} {nanocrystals}. \textit{Nano Lett.}
  \textbf{2020}, \textit{20}, 8626--8633, DOI:
  \doi{10.1021/acs.nanolett.0c03329}\relax
\mciteBstWouldAddEndPuncttrue
\mciteSetBstMidEndSepPunct{\mcitedefaultmidpunct}
{\mcitedefaultendpunct}{\mcitedefaultseppunct}\relax
\EndOfBibitem
\bibitem[Grigoryev \latin{et~al.}(2021)Grigoryev, Belykh, Yakovlev, Lhuillier,
  and Bayer]{grigoryev2021}
Grigoryev,~P.~S.; Belykh,~V.~V.; Yakovlev,~D.~R.; Lhuillier,~E.; Bayer,~M.
  Coherent {spin} {dynamics} of {electrons} and {holes} in {CsPbBr$_3$}
  {colloidal} {nanocrystals}. \textit{Nano Lett.} \textbf{2021}, \textit{21},
  8481--8487, DOI: \doi{10.1021/acs.nanolett.1c03292}\relax
\mciteBstWouldAddEndPuncttrue
\mciteSetBstMidEndSepPunct{\mcitedefaultmidpunct}
{\mcitedefaultendpunct}{\mcitedefaultseppunct}\relax
\EndOfBibitem
\bibitem[Kirstein \latin{et~al.}(2022)Kirstein, Kopteva, Yakovlev, Zhukov,
  Kolobkova, Kuznetsova, Belykh, Yugova, Glazov, Bayer, and
  Greilich]{kirstein2022SML}
Kirstein,~E.; Kopteva,~N.~E.; Yakovlev,~D.~R.; Zhukov,~E.~A.; Kolobkova,~E.~V.;
  Kuznetsova,~M.~S.; Belykh,~V.~V.; Yugova,~I.~A.; Glazov,~M.~M.; Bayer,~M.;
  Greilich,~A. Mode locking of hole spin coherences in {CsPb}({Cl},{Br})$_3$
  perovskite nanocrystals. \textbf{2022}; \url{http://arxiv.org/abs/2206.13323},
  arXiv:2206.13323 [cond-mat]\relax
\mciteBstWouldAddEndPuncttrue
\mciteSetBstMidEndSepPunct{\mcitedefaultmidpunct}
{\mcitedefaultendpunct}{\mcitedefaultseppunct}\relax
\EndOfBibitem
\bibitem[Tao \latin{et~al.}(2019)Tao, Schmidt, Brocks, Jiang, Tranca, Meerholz,
  and Olthof]{tao2019}
Tao,~S.; Schmidt,~I.; Brocks,~G.; Jiang,~J.; Tranca,~I.; Meerholz,~K.;
  Olthof,~S. Absolute energy level positions in tin- and lead-based halide
  perovskites. \textit{Nat. Commun.} \textbf{2019}, \textit{10}, 2560, DOI:
  \doi{10.1038/s41467-019-10468-7}\relax
\mciteBstWouldAddEndPuncttrue
\mciteSetBstMidEndSepPunct{\mcitedefaultmidpunct}
{\mcitedefaultendpunct}{\mcitedefaultseppunct}\relax
\EndOfBibitem
\bibitem[Baranowski \latin{et~al.}(2020)Baranowski, Plochocka, Plochocka, Su,
  Legrand, Barisien, Bernardot, Xiong, Testelin, and Chamarro]{baranowski2020b}
Baranowski,~M.; Plochocka,~P.; Plochocka,~P.; Su,~R.; Legrand,~L.;
  Barisien,~T.; Bernardot,~F.; Xiong,~Q.; Testelin,~C.; Chamarro,~M. Exciton
  binding energy and effective mass of {CsPbCl}$_{\textrm{3}}$: a
  magneto-optical study. \textit{Photonics Res.} \textbf{2020}, \textit{8},
  A50--A55, DOI: \doi{10.1364/PRJ.401872}\relax
\mciteBstWouldAddEndPuncttrue
\mciteSetBstMidEndSepPunct{\mcitedefaultmidpunct}
{\mcitedefaultendpunct}{\mcitedefaultseppunct}\relax
\EndOfBibitem
\bibitem[Straus and Kagan(2018)Straus, and Kagan]{straus2018}
Straus,~D.~B.; Kagan,~C.~R. Electrons, {excitons}, and {phonons} in
  {two}-{dimensional} {hybrid} {perovskites}: {connecting} {structural},
  {optical}, and {electronic} {properties}. \textit{J. Phys. Chem.
  Lett.} \textbf{2018}, \textit{9}, 1434--1447, DOI:
  \doi{10.1021/acs.jpclett.8b00201}\relax
\mciteBstWouldAddEndPuncttrue
\mciteSetBstMidEndSepPunct{\mcitedefaultmidpunct}
{\mcitedefaultendpunct}{\mcitedefaultseppunct}\relax
\EndOfBibitem
\bibitem[Stier \latin{et~al.}(2016)Stier, Wilson, Clark, Xu, and
  Crooker]{stier2016}
Stier,~A.~V.; Wilson,~N.~P.; Clark,~G.; Xu,~X.; Crooker,~S.~A. Probing the
  {influence} of {dielectric} {environment} on {excitons} in {monolayer}
  {WSe$_2$}: {insight} from {high} {magnetic} {fields}. \textit{Nano Lett.}
  \textbf{2016}, \textit{16}, 7054--7060, DOI:
  \doi{10.1021/acs.nanolett.6b03276}\relax
\mciteBstWouldAddEndPuncttrue
\mciteSetBstMidEndSepPunct{\mcitedefaultmidpunct}
{\mcitedefaultendpunct}{\mcitedefaultseppunct}\relax
\EndOfBibitem
\bibitem[Baranowski and Plochocka(2020)Baranowski, and
  Plochocka]{baranowski2020}
Baranowski,~M.; Plochocka,~P. Excitons in {metal}-{halide} {perovskites}.
  \textit{Adv. Energ. Mater.} \textbf{2020}, \textit{10}, 1903659, DOI:
  \doi{https://doi.org/10.1002/aenm.201903659}\relax
\mciteBstWouldAddEndPuncttrue
\mciteSetBstMidEndSepPunct{\mcitedefaultmidpunct}
{\mcitedefaultendpunct}{\mcitedefaultseppunct}\relax
\EndOfBibitem
\bibitem[Shornikova \latin{et~al.}(2021)Shornikova, Yakovlev, Gippius, Qiang,
  Dubertret, Khan, Di~Giacomo, Moreels, and Bayer]{shornikova2021}
Shornikova,~E.~V.; Yakovlev,~D.~R.; Gippius,~N.~A.; Qiang,~G.; Dubertret,~B.;
  Khan,~A.~H.; Di~Giacomo,~A.; Moreels,~I.; Bayer,~M. Exciton {binding}
  {energy} in {CdSe} {nanoplatelets} {measured} by {one}- and {two}-{photon}
  {absorption}. \textit{Nano Lett.} \textbf{2021}, \textit{21}, 10525--10531, DOI:
  \doi{10.1021/acs.nanolett.1c04159}\relax
\mciteBstWouldAddEndPuncttrue
\mciteSetBstMidEndSepPunct{\mcitedefaultmidpunct}
{\mcitedefaultendpunct}{\mcitedefaultseppunct}\relax
\EndOfBibitem
\bibitem[Neumann \latin{et~al.}(2021)Neumann, Feldmann, Moser, Delhomme,
  Zerhoch, van~de Goor, Wang, Dyksik, Winkler, Finley, Plochocka, Brandt,
  Faugeras, Stier, and Deschler]{neumann2021}
Neumann,~T.; Feldmann,~S.; Moser,~P.; Delhomme,~A.; Zerhoch,~J.; van~de
  Goor,~T.; Wang,~S.; Dyksik,~M.; Winkler,~T.; Finley,~J.~J.; Plochocka,~P.;
  Brandt,~M.~S.; Faugeras,~C.; Stier,~A.~V.; Deschler,~F. Manganese doping for
  enhanced magnetic brightening and circular polarization control of dark
  excitons in paramagnetic layered hybrid metal-halide perovskites. \textit{Nat.
  Commun.} \textbf{2021}, \textit{12}, 3489, DOI:
  \doi{10.1038/s41467-021-23602-1}\relax
\mciteBstWouldAddEndPuncttrue
\mciteSetBstMidEndSepPunct{\mcitedefaultmidpunct}
{\mcitedefaultendpunct}{\mcitedefaultseppunct}\relax
\EndOfBibitem
\bibitem[Fang \latin{et~al.}(2020)Fang, Yang, Adjokatse, Tekelenburg, Kamminga,
  Duim, Ye, Blake, Even, and Loi]{fang2020}
Fang,~H.-H.; Yang,~J.; Adjokatse,~S.; Tekelenburg,~E.; Kamminga,~M.~E.;
  Duim,~H.; Ye,~J.; Blake,~G.~R.; Even,~J.; Loi,~M.~A. Band-{edge} {exciton}
  {fine} {structure} and {exciton} {recombination} {dynamics} in {single}
  {crystals} of {layered} {hybrid} {perovskites}. \textit{Adv. Funct. Mater.}
  \textbf{2020}, \textit{30}, 1907979, DOI: \doi{10.1002/adfm.201907979}\relax
\mciteBstWouldAddEndPuncttrue
\mciteSetBstMidEndSepPunct{\mcitedefaultmidpunct}
{\mcitedefaultendpunct}{\mcitedefaultseppunct}\relax
\EndOfBibitem
\bibitem[Do \latin{et~al.}(2020)Do, Granados~del \'Aguila, Xing, Liu, and
  Xiong]{do2020}
Do,~T. T.~H.; Granados~del \'Aguila,~A.; Xing,~J.; Liu,~S.; Xiong,~Q. Direct
  and indirect exciton transitions in two-dimensional lead halide perovskite
  semiconductors. \textit{J. Chem. Phys.} \textbf{2020}, \textit{153}, 064705, DOI:
  \doi{10.1063/5.0012307}\relax
\mciteBstWouldAddEndPuncttrue
\mciteSetBstMidEndSepPunct{\mcitedefaultmidpunct}
{\mcitedefaultendpunct}{\mcitedefaultseppunct}\relax
\EndOfBibitem
\bibitem[Zhai \latin{et~al.}(2017)Zhai, Baniya, Zhang, Li, Haney, Sheng,
  Ehrenfreund, and Vardeny]{zhai2017}
Zhai,~Y.; Baniya,~S.; Zhang,~C.; Li,~J.; Haney,~P.; Sheng,~C.-X.;
  Ehrenfreund,~E.; Vardeny,~Z.~V. Giant {Rashba} splitting in {2D}
  organic-inorganic halide perovskites measured by transient spectroscopies.
  \textit{Sci. Adv.} \textbf{2017}, \textit{3}, e170070 DOI: \doi{10.1126/sciadv.1700704}\relax
\mciteBstWouldAddEndPuncttrue
\mciteSetBstMidEndSepPunct{\mcitedefaultmidpunct}
{\mcitedefaultendpunct}{\mcitedefaultseppunct}\relax
\EndOfBibitem
\bibitem[Kahmann \latin{et~al.}(2021)Kahmann, Duim, Fang, Dyksik, Adjokatse,
  Rivera~Medina, Pitaro, Plochocka, and Loi]{kahmann2021}
Kahmann,~S.; Duim,~H.; Fang,~H.-H.; Dyksik,~M.; Adjokatse,~S.;
  Rivera~Medina,~M.; Pitaro,~M.; Plochocka,~P.; Loi,~M.~A. Photophysics of
  {two}-{dimensional} {perovskites} -- {learning} from {metal} {halide}
  {substitution}. \textit{Adv. Funct. Mater.} \textbf{2021}, \textit{31}, 2103778,
  DOI: \doi{10.1002/adfm.202103778}\relax
\mciteBstWouldAddEndPuncttrue
\mciteSetBstMidEndSepPunct{\mcitedefaultmidpunct}
{\mcitedefaultendpunct}{\mcitedefaultseppunct}\relax
\EndOfBibitem
\bibitem[Liu \latin{et~al.}(2019)Liu, Sun, Gan, del \'Aguila, Fang, Xing, Do,
  White, Li, Huang, and Xiong]{liu2019b}
Liu,~S.; Sun,~S.; Gan,~C.~K.; del \'Aguila,~A.~G.; Fang,~Y.; Xing,~J.; Do,~T.
  T.~H.; White,~T.~J.; Li,~H.; Huang,~W.; Xiong,~Q. Manipulating efficient
  light emission in two-dimensional perovskite crystals by pressure-induced
  anisotropic deformation. \textit{Sci. Adv.} \textbf{2019}, \textit{5}, eaav9445,
  DOI: \doi{10.1126/sciadv.aav9445}\relax
\mciteBstWouldAddEndPuncttrue
\mciteSetBstMidEndSepPunct{\mcitedefaultmidpunct}
{\mcitedefaultendpunct}{\mcitedefaultseppunct}\relax
\EndOfBibitem
\bibitem[Yakovlev and Bayer(2017)Yakovlev, and Bayer]{yakovlevCh6}
Yakovlev,~D.~R.; Bayer,~M. Chapter 6 on \textit{Coherent Spin Dynamics of Carriers}. In \textit{Spin {Physics} in {Semiconductors}};
  Dyakonov,~M.~I., Ed.;   Springer International Publishing: Cham, 2017; pp 155--206, DOI:
  \doi{10.1007/978-3-319-65436-2_6}\relax
\mciteBstWouldAddEndPuncttrue
\mciteSetBstMidEndSepPunct{\mcitedefaultmidpunct}
{\mcitedefaultendpunct}{\mcitedefaultseppunct}\relax
\EndOfBibitem
\bibitem[Evers \latin{et~al.}(2021)Evers, Kopteva, Yugova, Yakovlev, Reuter,
  Wieck, Bayer, and Greilich]{evers2021}
Evers,~E.; Kopteva,~N.~E.; Yugova,~I.~A.; Yakovlev,~D.~R.; Reuter,~D.;
  Wieck,~A.~D.; Bayer,~M.; Greilich,~A. Suppression of nuclear spin
  fluctuations in an {InGaAs} quantum dot ensemble by {GHz}-pulsed optical
  excitation. \textit{npj Quantum Inf.} \textbf{2021}, \textit{7}, 1--7, DOI:
  \doi{10.1038/s41534-021-00395-1}\relax
\mciteBstWouldAddEndPuncttrue
\mciteSetBstMidEndSepPunct{\mcitedefaultmidpunct}
{\mcitedefaultendpunct}{\mcitedefaultseppunct}\relax
\EndOfBibitem
\bibitem[Dhanabalan \latin{et~al.}(2020)Dhanabalan, Leng, Biffi, Lin, Tan,
  Infante, Manna, Arciniegas, and Krahne]{dhanabalan2020}
Dhanabalan,~B.; Leng,~Y.-C.; Biffi,~G.; Lin,~M.-L.; Tan,~P.-H.; Infante,~I.;
  Manna,~L.; Arciniegas,~M.~P.; Krahne,~R. Directional {anisotropy} of the
  {vibrational} {modes} in {2D}-{layered} {perovskites}. \textit{ACS Nano}
  \textbf{2020}, \textit{14}, 4689--4697, DOI:
  \doi{10.1021/acsnano.0c00435}\relax
\mciteBstWouldAddEndPuncttrue
\mciteSetBstMidEndSepPunct{\mcitedefaultmidpunct}
{\mcitedefaultendpunct}{\mcitedefaultseppunct}\relax
\EndOfBibitem
\bibitem[Jacoby \latin{et~al.}(2022)Jacoby, Crane, and Gamelin]{jacoby2022}
Jacoby,~L.~M.; Crane,~M.~J.; Gamelin,~D.~R. Coherent {spin} {dynamics} in
  {vapor}-{deposited} {CsPbBr$_3$} {perovskite} {thin} {films}. \textit{Chem.
  Mater.} \textbf{2022}, \textit{34}, 1937--1945, DOI:
  \doi{10.1021/acs.chemmater.1c04382}\relax
\mciteBstWouldAddEndPuncttrue
\mciteSetBstMidEndSepPunct{\mcitedefaultmidpunct}
{\mcitedefaultendpunct}{\mcitedefaultseppunct}\relax
\EndOfBibitem
\bibitem[Heisterkamp \latin{et~al.}(2015)Heisterkamp, Zhukov, Greilich,
  Yakovlev, Korenev, Pawlis, and Bayer]{heisterkamp2015}
Heisterkamp,~F.; Zhukov,~E.~A.; Greilich,~A.; Yakovlev,~D.~R.; Korenev,~V.~L.;
  Pawlis,~A.; Bayer,~M. Longitudinal and transverse spin dynamics of
  donor-bound electrons in fluorine-doped {ZnSe}: {Spin} inertia versus {Hanle}
  effect. \textit{Phys. Rev. B} \textbf{2015}, \textit{91}, 235432, DOI:
  \doi{10.1103/PhysRevB.91.235432}\relax
\mciteBstWouldAddEndPuncttrue
\mciteSetBstMidEndSepPunct{\mcitedefaultmidpunct}
{\mcitedefaultendpunct}{\mcitedefaultseppunct}\relax
\EndOfBibitem
\bibitem[Smirnov \latin{et~al.}(2020)Smirnov, Zhukov, Yakovlev, Kirstein,
  Bayer, and Greilich]{smirnov2020}
Smirnov,~D.~S.; Zhukov,~E.~A.; Yakovlev,~D.~R.; Kirstein,~E.; Bayer,~M.;
  Greilich,~A. Spin polarization recovery and {Hanle} effect for charge
  carriers interacting with nuclear spins in semiconductors. \textit{Phys. Rev.
  B} \textbf{2020}, \textit{102}, 235413, DOI:
  \doi{10.1103/PhysRevB.102.235413}\relax
\mciteBstWouldAddEndPuncttrue
\mciteSetBstMidEndSepPunct{\mcitedefaultmidpunct}
{\mcitedefaultendpunct}{\mcitedefaultseppunct}\relax
\EndOfBibitem
\bibitem[Posmyk \latin{et~al.}(2022)Posmyk, Zawadzka, Dyksik, Surrente, Maude,
  Kazimierczuk, Babi\'nski, Molas, Paritmongkol, M\k{a}czka, Tisdale,
  P\l{}ochocka, and Baranowski]{posmyk2022}
Posmyk,~K.; Zawadzka,~N.; Dyksik,~M.; Surrente,~A.; Maude,~D.~K.;
  Kazimierczuk,~T.; Babi\'nski,~A.; Molas,~M.~R.; Paritmongkol,~W.;
  M\k{a}czka,~M.; Tisdale,~W.~A.; P\l{}ochocka,~P.; Baranowski,~M.
  Quantification of {exciton} {fine} {structure} {splitting} in a
  {two}-{dimensional} {perovskite} {compound}. \textit{J. Phys. Chem. Lett.}
  \textbf{2022}, \textit{13}, 4463--4469, DOI:
  \doi{10.1021/acs.jpclett.2c00942}\relax
\mciteBstWouldAddEndPuncttrue
\mciteSetBstMidEndSepPunct{\mcitedefaultmidpunct}
{\mcitedefaultendpunct}{\mcitedefaultseppunct}\relax
\EndOfBibitem
\bibitem[Quarti \latin{et~al.}(2020)Quarti, Katan, and Even]{quarti2020}
Quarti,~C.; Katan,~C.; Even,~J. Physical properties of bulk, defective, {2D}
  and {0D} metal halide perovskite semiconductors from a symmetry perspective.
  \textit{J. Phys. Materials} \textbf{2020}, \textit{3}, 042001, DOI:
  \doi{10.1088/2515-7639/aba6f6}\relax
\mciteBstWouldAddEndPuncttrue
\mciteSetBstMidEndSepPunct{\mcitedefaultmidpunct}
{\mcitedefaultendpunct}{\mcitedefaultseppunct}\relax
\EndOfBibitem
\end{mcitethebibliography}

\providecommand{\latin}[1]{#1}
\makeatletter
\providecommand{\doi}
  {\begingroup\let\do\@makeother\dospecials
  \catcode`\{=1 \catcode`\}=2 \doi@aux}
\providecommand{\doi@aux}[1]{\endgroup\texttt{#1}}
\makeatother
\providecommand*\mcitethebibliography{\thebibliography}
\csname @ifundefined\endcsname{endmcitethebibliography}
  {\let\endmcitethebibliography\endthebibliography}{}

\makeatletter
\setlength\acs@tocentry@height{0.45\textwidth}
\setlength\acs@tocentry@width{0.8\textwidth}
\makeatother

\begin{tocentry}

\includegraphics[width=\columnwidth]{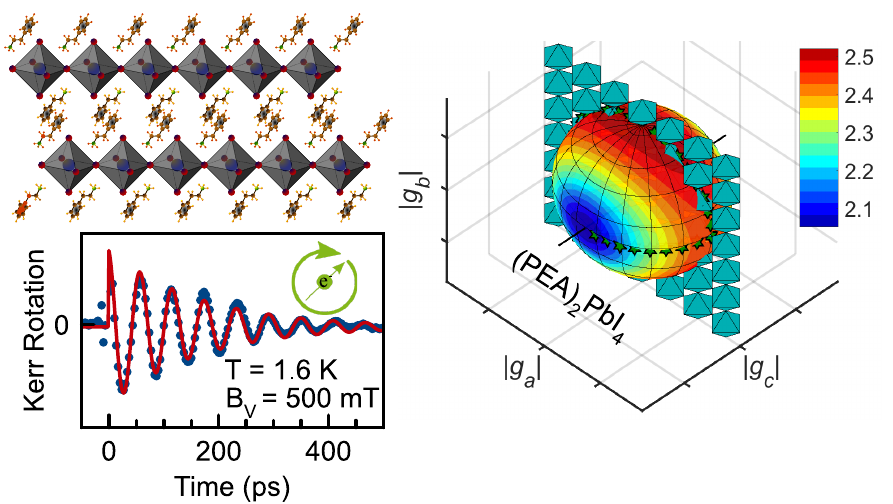}

\end{tocentry}
\end{document}